\documentstyle[aps,preprint,epsf]{revtex}
\begin{document}
\draft
\title{The Statistical Distributions of Level Widths and Conductance Peaks
       in Irregularly Shaped Quantum Dots}
\author{Y. Alhassid$^1$ and C. H. Lewenkopf$^2$}
\address{$^1$ Center for Theoretical Physics, Sloane Physics Laboratory,
	      Yale University, New Haven, CT 06520 USA\\
	 $^2$ Department of Physics, FM-15, University of Washington,
	      Seattle, WA 98195, USA}
\maketitle
\begin{abstract}
Analytical expressions for width and conductance peak distributions
for quantum dots with multi-channel leads in the Coulomb blockade
regime are presented for both limits of conserved and broken time-reversal
symmetry.
The results are valid for any number of non-equivalent
and correlated channels, and the distributions are expressed
in terms of the channel correlation matrix $M$ in each lead.
The matrix $M$ is also given in closed form.
A chaotic billiard is used as a model to test numerically the
theoretical predictions.\\
\end{abstract}

\narrowtext

 Advances in nanostructures technology make possible the manufacture of
semiconductor devices known as quantum dots \cite{Kas92} where electrons
are confined to very small two-dimensional regions.
By connecting external leads to such devices it is possible to
study their electronic transport properties: the conductance
can be measured as a function of the Fermi energy and/or as a
function of a magnetic flux through the dot.
Of particular interest are dots that are weakly coupled to leads,
due to the presence of barriers at the interface between dot and
leads.
In these cases the resonance widths $\Gamma$ are small compared to
their mean spacing $\Delta$.
At low temperatures $kT < \Delta$, only one quasi-bound level
participates in the conduction process.
The resulting Coulomb blockade peaks of the conductance \cite{MKW90}
are equally spaced but their amplitude exhibits order of magnitude
fluctuations.

For dot sizes that are smaller than the electron-impurity mean free
path, conductance fluctuations are determined by the
dot geometry.
We discuss dot geometries which display classical chaotic motion.
In such cases one can model their  transport properties using
the concepts of chaotic scattering \cite{LW91}.
Our results should also hold for weakly disordered dots
in the quasi-zero-dimension limit.
In Ref. \cite{JSA92} a statistical description of the conductance
peaks in the Coulomb blockade regime was developed using the
$R$-matrix formalism \cite{LT} and assuming that the dot's wavefunctions
 are described by random matrix theory (RMT).
Conductance and decay width distributions were derived for one-channel
leads, as well as for symmetric leads with several equivalent and
uncorrelated channels.
However, a more general  theory would encompass
leads with an arbitrary number of correlated and non-equivalent
channels.  Progress in this direction was made in
 Ref.\cite{MPA94}, where  by
modelling leads as point contacts \cite{PEI93}  spatial wave function
correlations were taken into account. Using the supersymmetry method
 \cite{Efe83}, exact formulas for the width and conductance peak
distributions were obtained for two-point contact leads.
However, the derivation was restricted  to the case of broken time
reversal symmetry, and could not be applied to leads with finite width
or  with more than two-point contacts.
  On the experimental side, we note that conductance distributions
 are becoming accessible; for
ballistic open dots
($\Gamma  \gg \Delta$) such distributions were recently measured \cite{CCM},
  and similar experiments are underway in the Coloumb blockade regime.

In this letter we present exact formulas for the width and conductance
peak distributions for leads with any number of channels that are in
general correlated and non-equivalent.
These distributions are obtained both for conserved and
broken time-reversal symmetry and are completely characterized in
terms of the channel correlation matrix $M$.
Our results are valid for both the point-like contacts and the continuous
extended leads models.
We are able to treat the most general case  because our methods are
based exclusively on RMT, which is technically simpler than the
supersymmetry method used in Ref. \cite{MPA94}.
To test our theory  we use a
chaotic billiard, the Africa \cite{Rob83}, for which the statistical
distributions of one-channel leads were recently studied in detail
\cite{BS94}.
This model is particularly useful to study systems with strong channel
correlations.
In contrast, the correlations between nearest points
in the discretized Anderson model of a disordered dot \cite{MPA94}
 were too weak to produce any
significant change in the width distribution of two-point leads (as
compared with the uncorrelated channels distribution). We remark that
since the partial width is analogous to the wavefunction intensity (see Eq.
(\ref{3}) below),
our results for the partial and total width distributions can be directly
 tested in the microwave cavity experiments \cite{AGH95},
where the wavefunctions intensities are measured at several points
and are spatially correlated.

Provided that $\Gamma \ll kT < \Delta$,  which is  typical of
experiments \cite{Kas92,MKW90}, the conductance peak amplitude for a two
lead geometry is given by \cite{Bee91}
\begin{eqnarray}
\label{1}
   G=\frac{e^2}{h}\, \frac{\pi}{2  kT}\, g \;,
\quad \mbox{where} \quad
   g=\frac{\Gamma^l_\lambda \Gamma_\lambda^r}{\Gamma_\lambda^l
 + \Gamma_\lambda^r} \;,
\end{eqnarray}
and $\Gamma_\lambda^l$ ($\Gamma_\lambda^r$) is the partial decay width
of a resonance $\lambda$ into channels of the left (right) lead.
Each lead can support $\Lambda^{l(r)}$ open channels  so that
$\Gamma_\lambda^{l(r)} = \sum_c |\gamma_{c\lambda}^{l(r)}|^2$, where
$\gamma_{c\lambda}^{l(r)}$ is the partial amplitude to decay into channel
$c$ on the left (right).
$R$-matrix theory gives \cite{LT}
\begin{eqnarray}
\label{2}
\gamma_{c\lambda} = \left( \frac{\hbar^2 k_cP_c}{m} \right)^{1/2}
      \int \! dS\, \Phi_c^\ast({\bf r}) \Psi_\lambda({\bf r})  \;,
\end{eqnarray}
where $\Psi_\lambda$ is the resonance wave function inside the dot (scattering
region), $\Phi_c$ is the wave function of an open channel $c$ in the lead
(asymptotic region), and the integral is taken over the contact boundary
between
the lead and the dot.  $P_c$ and $k_c$  are the channel penetration
 factor to tunnel through the barrier and the longitudinal wave number,
respectively.
An alternative modelling assumes that the quantum dot is connected to the
leads by one or more point-like contacts \cite{PEI93}.
Every such point contact ${\bf r}_c$ is considered as a channel
and the corresponding  partial amplitude is \cite{MPA94}
\begin{eqnarray}
\label{3}
 \gamma_{c\lambda} =
 \left( {\alpha_c\cal A} \Delta/\pi \right)^{1/2} \Psi_\lambda({\bf r}_c) \;,
\end{eqnarray}
where ${\cal A}$ is the area of the dot, $\Delta$ is the mean
resonance spacing and $\alpha_c$ is the coupling parameter of the point
contact to the dot . Expanding a resonance wave
function with energy $\varepsilon$ in a fixed basis of states
with that energy  $\Psi_\lambda({\bf r})= \sum_\lambda \psi_{\lambda \mu}
\rho_\mu({\bf r})$ (the sum is truncated to $N$ terms, typically
much larger than $\Lambda$),  the partial width to decay to channel
 $c$ can be expressed as a scalar product
$\gamma_{c\lambda} \equiv \langle \mbox{\boldmath$\phi$}_c
| \mbox{\boldmath$\psi$}_\lambda \rangle
         = \sum_\mu \phi_{c \mu}^* \psi_{\lambda\mu}$,
where $\phi_{c\mu}^* \equiv \left(\hbar^2 k_c P_c/m \right)^{1/2}\!\int \! dS
\,\Phi_c^*({\bf r}) \rho_\mu({\bf r})$ in the $R$-matrix formalism
 and  $\phi_{c\mu}^* \equiv \left( {\alpha_c\cal A} \Delta/\pi\right)^{1/2}
 \rho^*_\mu ({\bf r}_c)$ in the point-like contact model.
Expressing the width as a scalar product allows us to treat the extended
lead and the point-like contact models in an equivalent manner.

The resonance states $\Psi_\lambda$ are assumed to
have GOE- or GUE-like statistical properties, depending
on the symmetry class to which the dynamics in the dot corresponds
\cite{JSA92}.
This assumption is valid both for dots with chaotic dynamics
and for weakly disordered dots.
The eigenvector components  $ (\psi_1, \psi_2, \ldots, \psi_N)
\equiv \mbox{\boldmath$\psi$}$ (in the following we shall omit
the eigenvector label $\lambda$) are therefore randomly distributed
$P(\mbox{\boldmath$\psi$}) \propto \delta
(\sum_{\mu=1}^N |\psi_\mu|^2 - 1)$ \cite{BFF81}.
The joint distribution of the partial width  amplitudes
$\mbox{\boldmath$\gamma$} =(\gamma_1, \gamma_2,  \ldots,
\gamma_\Lambda)$ for $\Lambda$ channels is given by
\begin{eqnarray}
\label{5}
  P(\mbox{\boldmath$\gamma$}) \propto \! \int \! \!
    D[\mbox{\boldmath$\psi$}] \,
    \delta \! \left( \sum_{\mu=1}^N | \psi_\mu |^2 -1 \right)
    \prod_{c=1}^{\Lambda}{\delta (\gamma_c -
        \langle\mbox{\boldmath$\phi$}_c|\mbox{\boldmath$\psi$}\rangle)} \; ,
\end{eqnarray}
where the metric is
$D[\mbox{\boldmath$\psi$}] \equiv \prod_{\mu=1}^{N}{d\psi_\mu}$ for
the GOE and
$D[\mbox{\boldmath$\psi$}] \equiv \prod_{\mu=1}^{N}{d\psi^*_\mu
d\psi_\mu/2\pi i}$ for the GUE.
To evaluate (\ref{5}) we transform $\mbox{\boldmath$\phi$}_c =\sum_{c^\prime}
\hat{\mbox{\boldmath$\phi$}}_{c^\prime} F_{c^\prime c}$ to obtain a
new set of orthonormal channels $\langle \hat{\mbox{\boldmath$\phi$}}_c
|\hat{\mbox{\boldmath$\phi$}}_{c^\prime}
\rangle= \delta_{cc^\prime}$.
In the limit $N\rightarrow \infty$, provided that $\Lambda \ll N$, one
finds \cite{KP63}
\begin{eqnarray}
\label{6}
 P(\mbox{\boldmath$\gamma$}) = (\det M)^{-\beta/2}
      \mbox{e}^{- \frac{\beta}{2} \mbox{\boldmath$\gamma$}^\dagger
      M^{-1} \mbox{\boldmath$\gamma$}}\;,
\end{eqnarray}
where $M \equiv (NF^\dagger F)^{-1}$.
The distribution (\ref{6}) is normalized with the measure
$D[\mbox{\boldmath$\gamma$}] \equiv \prod_{c=1}^{\Lambda}
{d\gamma_c/2\pi}$ for the GOE ($\beta=1$) and
$D[\mbox{\boldmath$\gamma$}] \equiv \prod_{c=1}^{\Lambda}
{d\gamma^*_c d\gamma_c/2\pi i}$ for the GUE ($\beta=2$).
Note that for both ensembles the joint partial width amplitudes
distribution is Gaussian, and it follows that $M$ is just the correlation
 matrix of the partial widths
$M_{cc^\prime} = \overline{ \gamma_c^* \gamma_{c^\prime}} =
 \frac{1}{N} \langle \mbox{\boldmath$\phi$}_c
|\mbox{\boldmath$\phi$}_{c^\prime}\rangle$.
In general the channels are correlated (non-orthogonal)  and
non-equivalent, {\sl i.e.} have different average partial widths.

Recalling Eqs. (\ref{2}) and (\ref{3}), the spatial autocorrelation
function $C(\Delta{\bf r}) \equiv \overline{\Psi^*({\bf r})
\Psi({\bf r} + \Delta{\bf r})}/\overline{|\Psi({\bf r})|^2}$  plays
a central role in deriving explicit expressions for the correlation
matrix $M$.
For fully chaotic systems with time reversal symmetry, Berry
obtained $C(\Delta{\bf r})$ semiclassically, assuming that classical
orbits cover uniformly the energy surface  \cite{Ber77}.
For an eigenstate of a chaotic billiard with energy $\varepsilon =
\hbar^2 k^2/2m$, $C(\Delta{\bf r}) = J_0(k|\Delta{\bf r}|)$.
Alternatively, a resonance at energy $\varepsilon$ can be expanded
inside the dot in the fixed basis $\rho_\mu({\bf r})={\cal A}^{-1/2}
\exp(i{\bf k}_\mu \cdot {\bf r})$, where the $\mu$'s
correspond to $N$ different (random) orientations of ${\bf k}$.
In the spirit of RMT, assuming that the expansion coefficients
$\psi_\mu$
are Gaussian  (this hypothesis was confirmed for
the Africa billiard \cite{LR94}) and using
$\overline{\psi_\mu^\ast \psi_{\mu^\prime}} = N^{-1}\delta_{\mu \mu^\prime}$,
 one can  rederive Berry's result.
See also Ref. \cite{Pr95}.
The presence of a small magnetic flux $\Phi$ introduces corrections to
$C(\Delta{\bf r})$ which are small in the semiclassical limit
and are of order $\hbar^2(\Phi/\Phi_0)^2/(2m{\cal A}\varepsilon) \ll 1$.
Thus, the correlation matrix is given by
\begin{eqnarray}
\label{7}
 M_{cc^\prime} = \frac{\hbar^2
  (k_cP_c k_{c^\prime} P_{c^\prime})^{1/2}}{m {\cal A}}
         \int \!\! dS\!\int \!\! dS^\prime \, \Phi^\ast_c({\bf r})
              J_0(k|{\bf r} - {\bf r^\prime}|) \Phi_c({\bf r^\prime})
\end{eqnarray}
for the finite width leads, and
\begin{eqnarray}\label{8}
 M_{cc^\prime} = \frac{\Delta (\alpha_c \alpha_{c^\prime})^{1/2}}{\pi}
J_0(k|{\bf r}_c - {\bf r}_c^\prime|)
\end{eqnarray}
for the point contacts model.

We turn next to the calculation of the total width distribution
$P(\Gamma)$ in a given lead that supports $\Lambda$ channels and
is characterized by a correlation matrix $M$.
Using (\ref{6}) and $\Gamma= \sum_c |\gamma_c|^2$,
the characteristic function of $P(\Gamma)$ is readily obtained and we find

\begin{eqnarray}
\label{9}
 P(\Gamma) = \frac{1}{2 \pi} \int^{\infty}_{-\infty}
	     d t \,\frac{e^{-i t \Gamma}}{\left[\det (I - 2 i t M/\beta)
\right]^{\beta/2}} \;.
\end{eqnarray}
The matrix $M$ is hermitian and positive definite, so that its
eigenvalues $w_c^2$ are all positive.
Since $\Gamma$ is invariant under orthogonal (unitary) transformations,
$P(\Gamma)$ depends only on $w_c^2$ and Eq. (\ref{9}) can be evaluated
by contour integration.
When all eigenvalues of $M$ are non-degenerate, we find for the GUE case
\begin{eqnarray}
\label{10}
P_{GUE}(\Gamma) = \frac{1}{\prod_c w_c}
       \sum_{c=1}^\Lambda
       \mbox{e}^{-\Gamma/w_c^2}
       \left[\prod_{c^\prime \neq c}
       (\frac{1}{w_{c^\prime}^2} - \frac{1}{w_c^2})\right]^{-1} \!\!\! .
\end{eqnarray}
For two equivalent channels ($M_{11} = M_{22} = \overline{\Gamma}/2$),
Eq. (\ref{10}) coincides with the result of \cite{MPA94}
\begin{eqnarray}
\label{11}
 P_{GUE}({\hat\Gamma}) = \frac{2}{|f|}
       \mbox{e}^{-2 {\hat\Gamma}/(1-|f|^2) }
 \sinh \left( \frac{2|f|}{1 - |f|^2 }{\hat\Gamma} \right)\;,
\end{eqnarray}
where ${\hat\Gamma} = \Gamma/\overline{\Gamma}$, and
$f=M_{12}/\sqrt{M_{11}M_{22}}$ measures the degree of correlation
between the two channels.

For the GOE case, the integral in (\ref{9}) is still straightforward
but cannot be expressed in a simple form as above.
Labelling the inverse eigenvalues of $M$ in increasing order
$w_1^{-2} < w_2^{-2} < \ldots$, the contour integral gives
\begin{eqnarray}
\label{12}
 P_{GOE}(\Gamma) && =
     \left(\pi 2^{\Lambda/2} \prod_c  w_c\right)^{-1}
     \sum_{m=1} \int_{(2 w^{2}_{2m-1})^{-1}}^{(2 w^{2}_{2m})^{-1}}
     d \tau \times \nonumber \\
 && \times \frac{\mbox{e}^{-\Gamma \tau}}{\sqrt{
     \prod_{s=1}^{2m-1}    (\tau - \frac{1}{2 w^2_s})
     \prod_{s^\prime=2m}^\Lambda (\frac{1}{2 w^2_{s^\prime}} - \tau)}} \;,
\end{eqnarray}
where, for an odd number of channels $\Lambda$, we define
$1/2 w^2_{\Lambda+1} \rightarrow \infty$.
For the case of two equivalent but correlated channels, Eq. (\ref{12})
reduces to
\begin{eqnarray}
\label{13}
 P_{GOE}({\hat\Gamma}) =\frac{1}{\sqrt{1-f^2}}\,
                  \mbox{e}^ {-{\hat\Gamma}/(1-f^2)}\,
                  I_0\!\left( \frac{f}{1-f^2} {\hat\Gamma}\right) \;,
\end{eqnarray}
where $f=M_{12}/\sqrt{M_{11}M_{22}}$ and $I_0$ is the modified
Bessel function of order zero.

To test the RMT predictions, we modelled a quantum dot by the Africa
billiard \cite{Rob83,BS94}.
The shape of this billiard is determined by the image of the circle
of radius one in the complex $z$-plane under the conformal mapping
$w(z) = (z + bz^2 + ce^{i\delta}z^3)/\sqrt{1 + 2b^2 + 3c^2}$.
We studied the case $b=0.2,c=0.2$ and $\delta=\pi/2$ which has a classical
fully chaotic phase space \cite{Rob83}.
For the statistical study of the case of broken time reversal symmetry,
we consider the billiard threaded by an Aharonov-Bohm
flux line $\Phi =\alpha \Phi_0$ \cite{BR86,BS94}.
We choose $\alpha=1/4$ where the resonance fluctuations are known to
be GUE-like \cite{BS94}.
The study considers 50 eigenstates starting from the 100{\sl th}.

To investigate the eigenfunction amplitude correlations
$C(\Delta {\bf r})$  we have used the eigenfunctions
of the billiard with von Neumann boundary conditions. The results are
shown in the inset to Fig.\ \ref{1} where the correlations in the model
(solid line) compare well with the theoretical  result
$J_0(k\Delta r)$ (dashed line).  In what follows, we assume for simplicity
 that the penetration factors $k_c P_c$ are channel-independent.
We first studied extended leads by taking the contact region of the
lead and the dot  to have a finite width $D$.
In this case, the channels are defined by the allowed transverse
momenta $\pi c/D \;  (c= 1,\ldots,\Lambda)$ where $\Lambda =
\mbox{int}[kD/\pi]$.
To guarantee that the correlation matrix $M$ is the same for
different eigenfunctions of the billiard, we choose $D$ such
that $kD$ is constant and scale the partial amplitude (\ref{2})
by $k$. We find that the channels in this case are weakly correlated.
Thus, the total width distribution is similar to the
case of uncorrelated equivalent channels, which gives a
$\chi^2$ distribution with $\Lambda$ (GOE) or $2\Lambda$ (GUE)
degrees of freedom.
However this changes if the barrier penetration factors have
 a strong energy dependence.
Our model calculations agree nicely with the RMT predictions (see
Fig.\  \ref{1}).

We also studied the model of leads with $\Lambda$ point-like contacts
by choosing for each lead a sequence of $\Lambda$ equally spaced points
on the Africa boundary.
Thus, according to (\ref{8}), $M$ is completely determined by
$\Lambda$ and $k |\Delta {\bf r}|$ (where $|\Delta {\bf r}|$ is
the distance between two neighboring points).
In Fig. \ref{2} we compare the results of this model (histograms)
with the theoretical predictions (solid lines) for $\Lambda = 4$
and different values of $k |\Delta {\bf r}|$.

To calculate the conductance distribution, which is the measurable quantity
for quantum dots, we assume
that the left and right leads are uncorrelated
and characterized by $M^l$ and $M^r$, respectively.
We then use $P(g) = \int d\Gamma^l d\Gamma^r \delta \! \left( g -
\Gamma^l \Gamma^r/(\Gamma^l + \Gamma^r)\right) P(\Gamma^l)
P(\Gamma^r)$, where
$P(\Gamma)$ is given by (\ref{10}) or (\ref{12}).
In the absence of  time reversal symmetry we find
\begin{eqnarray}
\label{14}
 P&&_{GUE}(g) =  \frac{16 g}{\left(\prod_c v_c \prod_d w_d\right)^{2}}
  \sum_{c,d} \mbox{e}^{-(\frac{1}{v_c^{2}} + \frac{1}{w_d^{2}})g}
 \nonumber \\
 && \times \left[
  \prod_{c^\prime \neq c} (\frac{1}{v_{c^\prime}^{2}} - \frac{1}{v_c^{2}})
  \prod_{d^\prime \neq d} (\frac{1}{w_{d^\prime}^{2}} - \frac{1}{w_d^{2}})
    \right]^{-1}
 \nonumber \\
 && \times \left[ K_0\left(\frac{2g}{v_c w_d}\right)
 + \frac{1}{2}\left(\frac{ v_c }{ w_d}+\frac{ w_d}{ v_c} \right)
 K_1\left( \frac{2g}{v_c  w_d} \right) \right] \;,
\end{eqnarray}
where $v_c$ ($w_d$) are the eigenvalues of $M^l$ ($M^r$) and $K_0$
($K_1$) is the modified  Bessel functions of order zero (one).

The result of\cite{JSA92,PEI93} is a special case of (\ref{14}) for
one channel leads with $\overline{\Gamma}^l = \overline{\Gamma}^r$
({\sl i.e.}  $v_1=w_1$), while the distribution of Ref. \cite{MPA94}
is obtained for two (equivalent) channels leads.
For time reversal symmetric systems, we also obtained a closed formula
for $P(g)$, which has similar structure to (\ref{14}).
Fig. \ref{3} shows a comparison between the theoretical conductance
distributions and those calculated for the Africa billiard for
$\Lambda$-point symmetric leads with $k |\Delta {\bf r}| =1$ and for
various values of $\Lambda$.

In conclusion, we have derived in closed form the width and conductance
peak distributions in a quantum dot, for leads with any number of
correlated and/or non-equivalent channels, and in the presence or
absence of time reversal symmetry.
The only required input to determine the distributions is the channel
correlation matrix $M$, for which an explicit expression was obtained.
Our results for the decay widths could also be  applied
 to compound nucleus
reactions in the limit of isolated resonances, where $M$ is
evaluated by the optical model.

We acknowledge A.D. Stone and H. Bruus for useful discussions
 and for the use of their Africa
billiard computer program and E. R. Mucciolo for discussions.
This work was supported in part by  DOE Grant
DE-FG02-91ER40608 and by NSF.
C.H.L. thanks the Center for Theoretical Physics at
Yale University for its hospitality.

%
%

\begin{figure}

\vspace{3 mm}

\caption
{Total width distributions $P(\Gamma)$ for extended leads
with $\Lambda = 4$.
$\Gamma$ is measured in units of its average value.
The solid lines are the theoretical distributions and the
histograms are results from the Africa billiard
for the (a) GOE and
(b) GUE cases (see text).
The insert shows the spatial wavefunction correlation
$C(\Delta {\bf r})$ calculated for the Africa (solid
line) compared with $J_0 (k|\Delta {\bf r}|)$ (dashed
line).}
\label{fig 1}

\vspace{10 mm}

\caption
{Total width distributions $P(\Gamma)$ for $\Lambda=4$ point-contact leads.
(a) GOE, $k|\Delta {\bf r}| = 0.5$;
(b) GUE, $k|\Delta {\bf r}| = 0.5$;
(c) GOE, $k|\Delta {\bf r}| = 1$ and
(d) GUE, $k|\Delta {\bf r}| = 1$.
The solid lines correspond to the theoretical distributions,
the dashed lines to uncorrelated channels and the histograms
to the Africa billiard calculations.}
\label{fig2}

\vspace{10 mm}

\caption
{Conductance peak distributions $P(g)$ for $\Lambda$ point-contact
symmetric leads ($M^l=M^r$) with $k|\Delta {\bf r}|=1$.
(a) GOE, $\Lambda=2$;
(b) GUE, $\Lambda=2$;
(c) GOE, $\Lambda=4$ and
(d) GUE, $\Lambda=4$.
The convention for the lines is as in Fig. 2 .}
\label{fig3}
\end{figure}

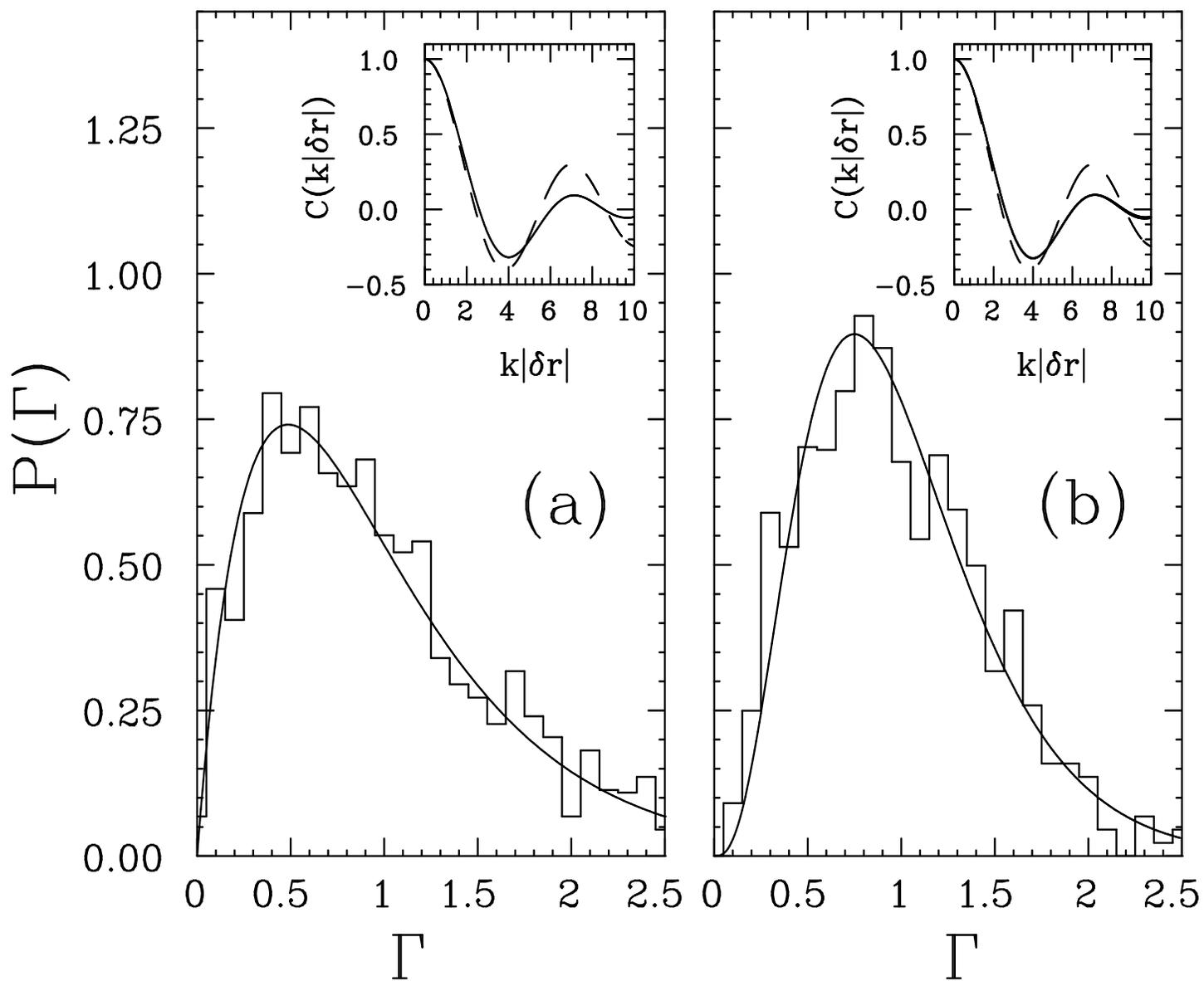
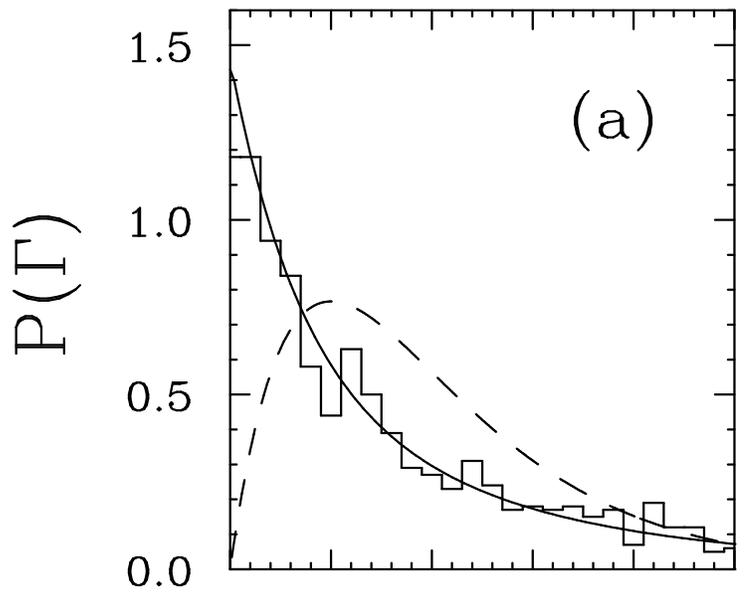
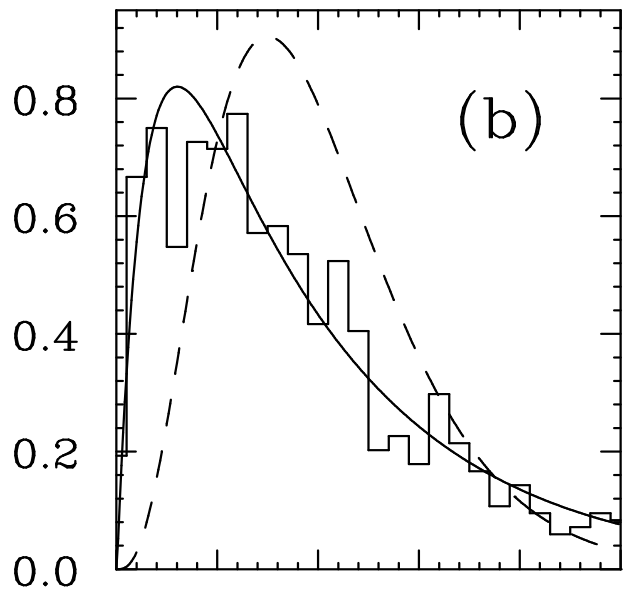
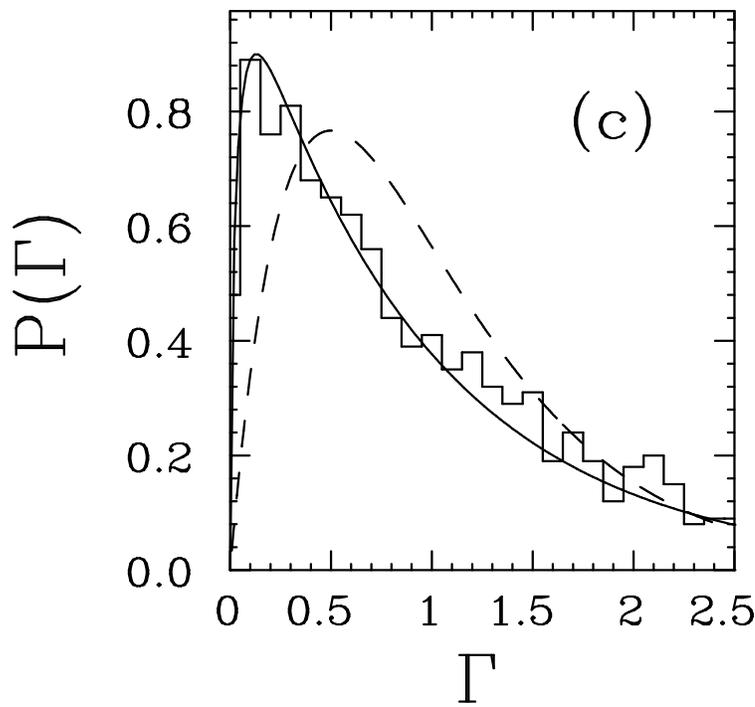
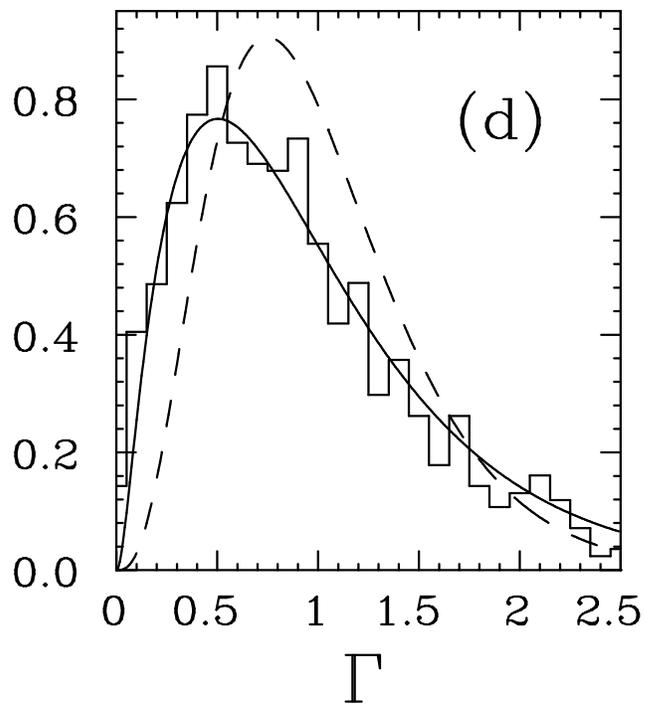
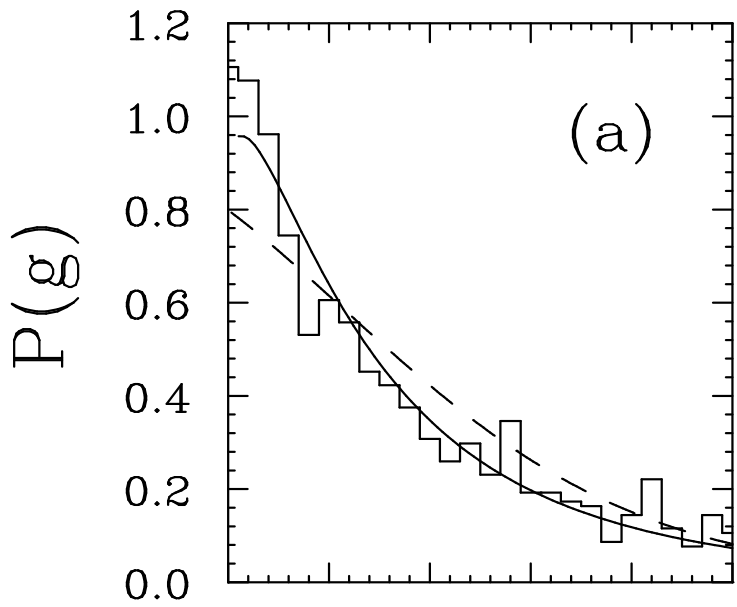
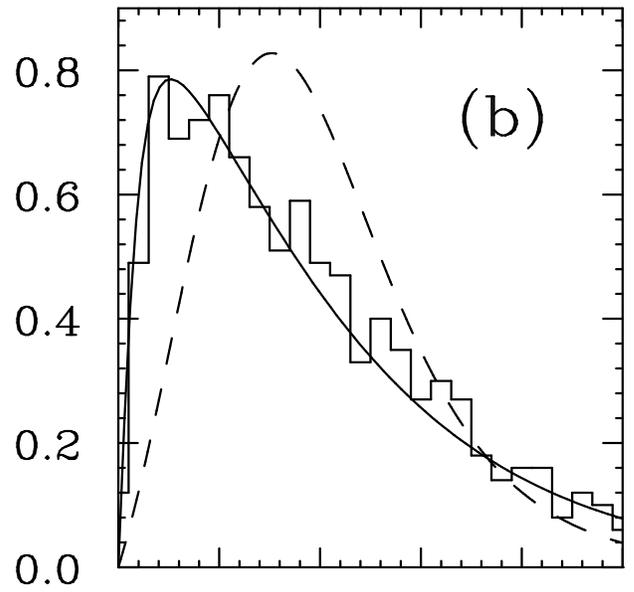
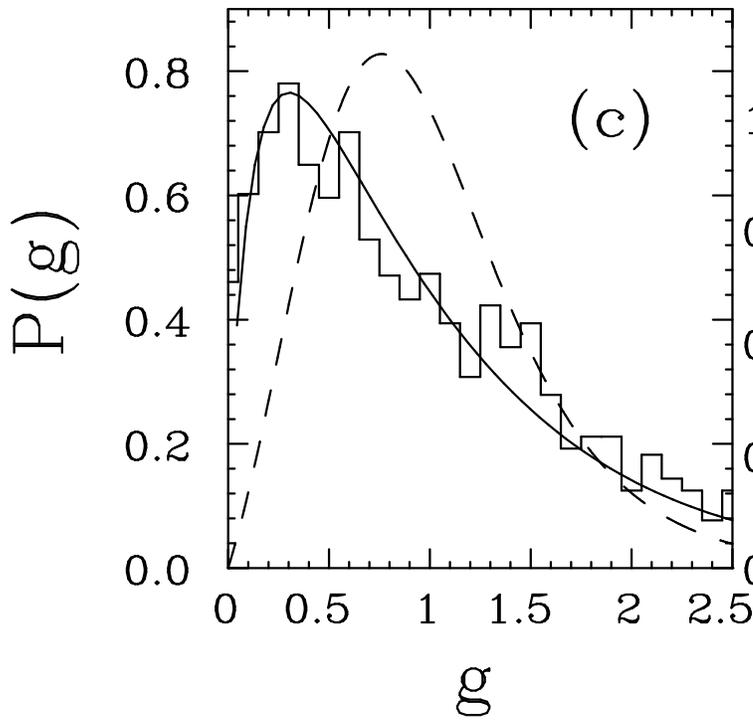
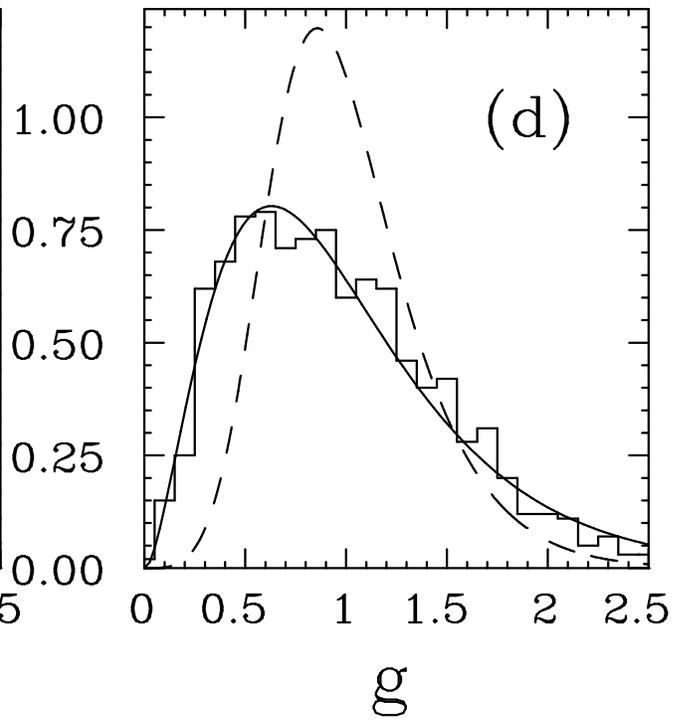
\end{document}